\documentclass[twocolumn]{aastex631}
\usepackage{hyperref}
\hypersetup{
  linkcolor=red,
  citecolor=blue,
  filecolor=cyan,
  urlcolor=magenta
}
\usepackage{amsmath}
\usepackage{graphicx}
\usepackage{placeins}
\usepackage{booktabs}
\usepackage{gensymb}
\usepackage{upgreek}
\usepackage{multirow}
\usepackage{soul}
\usepackage[normalem]{ulem}
\usepackage{xcolor}
\usepackage[utf8]{inputenc} 
\usepackage[T1]{fontenc}
\usepackage{textcomp}

\newcommand{\kms}{km~s$^{-1}$}

\newcommand{\msun}{$M_{\mathrm{\odot}}$}
\newcommand{\MHI}{$M_{\rm H\, \textsc{i}}$}
\newcommand{\mstar}{$M_{*}$}
\def\HI{H\,  {\textsc{\romannumeral 1}}}

\received{2026.01.27}
\revised{xxxx}
\accepted{xxxx}
\submitjournal{ApJ}

\begin{document}
\title{Insights into the Physical Nature of Polar Ring Galaxies from H I Observations}

\correspondingauthor{Niankun Yu}
\email{niankunyu@bao.ac.cn}

\shorttitle{TFR of PRGs}
\shortauthors{Yu et~al. 2025}

\author[0000-0002-9066-370X]{Niankun Yu}
\affiliation{National Astronomical Observatories, Chinese Academy of Sciences, 20A Datun Road, Beijing 100101, People's Republic of China}
\affiliation{State Key Laboratory of Radio Astronomy and Technology, National Astronomical Observatories, Chinese Academy of Sciences, Beijing 100101, People's Republic of China}
\affiliation{Max Planck Institute for Radio Astronomy, Auf dem H$\ddot{u}$gel 69, 53121 Bonn, Germany}

\author{Han Zheng}
\affiliation{S.K.Lee Honors College, China University of Geosciences, Wuhan, 430074,  People's Republic of China}

\author[0000-0002-9390-9672]{Chao-Wei Tsai}
\affiliation{National Astronomical Observatories, Chinese Academy of Sciences, 20A Datun Road, Beijing 100101,  
People's Republic of China}
\affiliation{Institute for Frontiers in Astronomy and Astrophysics, Beijing Normal University, Beijing 102206, China}
\affiliation{School of Astronomy and Space Science, University of Chinese Academy of Sciences, Beijing 100049,  
People's Republic of China}
\affiliation{State Key Laboratory of Radio Astronomy and Technology, National Astronomical Observatories, Chinese Academy of Sciences, Beijing 100101, People's Republic of China}

\author[0000-0003-3948-9192]{Pei Zuo}
\affiliation{National Astronomical Observatories, Chinese Academy of Sciences, 20A Datun Road, Beijing 100101,  
People's Republic of China}
\affiliation{State Key Laboratory of Radio Astronomy and Technology, National Astronomical Observatories, Chinese Academy of Sciences, Beijing 100101, People's Republic of China}

\author[0000-0001-6947-5846]{Luis C. Ho}
\affiliation{Kavli Institute for Astronomy and Astrophysics, Peking University, Beijing 100871, People's Republic of China}
\affiliation{Department of Astronomy, School of Physics, Peking University, Beijing 100871, People's Republic of China}

\author[0000-0003-4357-3450]{Am$\'{e}$lie Saintonge}
\affiliation{Max Planck Institute for Radio Astronomy, Auf dem H$\ddot{u}$gel 69, 53121 Bonn, Germany}

\author{Zheng Zheng}
\affiliation{Department of Physics, Anhui Normal University, Wuhu, Anhui 241002, People's Republic of China}
\affiliation{National Astronomical Observatories, Chinese Academy of Sciences, Beijing 100101, People's Republic of China}
\affiliation{State Key Laboratory of Radio Astronomy and Technology, National Astronomical Observatories, Chinese Academy of Sciences, Beijing 100101, People's Republic of China}

\author[0000-0003-3523-7633]{Nathan Deg}
\affiliation{Department of Physics, Engineering Physics, and Astronomy, Queen's University, Kingston ON K7L 3N6, Canada}

\author[0000-0002-2169-0472]{Ningyu Tang}
\affiliation{Department of Physics, Anhui Normal University, Wuhu, Anhui 241002, People's Republic of China}

\author{Xin Ai}
\affiliation{School of Physics and Astronomy, Yunnan University, Kunming, 650091, People's Republic of China}

\author[0000-0001-6106-1171]{Junzhi Wang}
\affiliation{School of Physical Science and Technology, Guangxi University, Nanning 530004, People's Republic of China}

\author{Xiang Jie}
\affiliation{College of Mathematics and Physics, Shanghai Normal University, Shanghai, 200234, People's Republic of China}

\author[0000-0003-3010-7661]{Di Li}
\affiliation{New Cornerstone Science Laboratory, Department of Astronomy, Tsinghua University, Beijing 100084, People's Republic of China}
\affiliation{National Astronomical Observatories, Chinese Academy of Sciences, Beijing 100101, People's Republic of China}

\begin{abstract}
Polar ring galaxies (PRGs) host an outer ring of gas and stars oriented nearly perpendicular to the main stellar body. They represent extreme examples of misaligned systems and provide valuable insight into galaxy interactions, gas accretion, and peculiar gas dynamics. We compile a complete sample of kinematically confirmed PRGs and collect their \HI\ measurements. Combining literature data with new observations from FAST, we detect \HI\ emission in 22 sources, identify one potential \HI\ absorption feature, and find four non-detections among 40 confirmed PRGs. Compared to galaxies in the ALFALFA and xGASS surveys, PRGs predominantly occupy the green valley or quenched regimes but exhibit higher gas fractions than typical early-type galaxies, suggesting gas accretion. The \HI\ profile asymmetry and shape for PRGs are not consistent with that of the ALFALFA sample with $p<$0.05. We examine their Tully$-$Fisher relation (TFR) and baryonic TFR (bTFR), linking the systems' rotation velocities to their masses. The extreme outliers in TFRs for the control sample tend to display single-peaked \HI\ profiles. PRGs do not follow a tight TFR or bTFR if the H I resides primarily in the host galaxy. But the scatter decreases significantly if we assume the gas is mainly distributed in the polar ring. Spatially resolved \HI\ observations are essential to disentangle the gas distribution and kinematics in PRGs, which are key to understanding their formation mechanisms.
\end{abstract}
\keywords{Galaxies: fundamental parameters - galaxies: ISM - galaxies: polar ring - radio lines, \HI\ 21~cm}
\section{Introduction}
\label{sec:intro}
Polar ring galaxies (PRGs) consist of a central host galaxy surrounded by an outer ring of gas, dust, and stars \citep{Schweizer1983AJ.....88..909S, Varnas1987ApJ...313...69V}. The ring rotates in a plane nearly perpendicular to that of the host, providing a distinctive configuration for exploring galactic structure and dynamics \citep{Whitmore1990AJ....100.1489W, BinneyMerrifield1998gaas.book.....B}. Early identifications of PRGs, such as the peculiar NGC~2685, revealed striking perpendicular rings and unusual kinematic behavior within the optical plane \citep{Sandage1961hag..book.....S, Schechter1978AJ.....83.1360S}. These studies established the foundation for understanding the kinematic misalignment between host galaxies and their polar rings.

Building upon these early discoveries, \citet{Whitmore1990AJ....100.1489W} compiled a comprehensive catalog of confirmed and candidate PRGs, including six kinematically verified systems, 27 good candidates, 73 possible candidates, and 51 related objects. They estimated that roughly 0.5\% of nearby lenticular (S0) galaxies currently host polar rings, while up to 5\% may have done so at some stage of their evolution after correcting for selection effects. Subsequent efforts based on the Sloan Digital Sky Survey (SDSS; \citealt{York2000AJ....120.1579Y}) and the Galaxy Zoo project \citep{Lintott2008MNRAS.389.1179L} expanded the catalog to 70 best candidates, 115 good candidates, 53 PRG-related systems, and 37 galaxies exhibiting nearly face-on polar rings \citep{Moiseev2011MNRAS.418..244M}. Follow-up spectroscopy confirmed ten of these as kinematically verified PRGs, such as \citet{Moiseev2011MNRAS.418..244M, Moiseev2014ASPC..486...71M}. Despite these advances, confirmed PRGs remain rare, limiting a detailed understanding of their formation and evolutionary pathways.

The formation and longevity of polar rings have been attributed to several mechanisms: (1) tidal interactions or gas accretion from nearby companions \citep{Bournaud2003A&A...401..817B}, and (2) accretion of pristine gas from the intergalactic medium (IGM; \citealt{Whitmore1990AJ....100.1489W, Spavone2010ApJ...714.1081S, Moiseev2011MNRAS.418..244M}). Because the rings are dynamically stable over a few gigayears and contain abundant neutral atomic hydrogen (\HI\ 21~cm; \citealt{Richter1994AJ....107...99R}), a triaxial dark matter halo is likely required to maintain their stability \citep{Moiseev2015BaltA..24...76M}. The polar rings are typically younger and more gas-rich than their host galaxies \citep{Reshetnikov1997A&A...325..933R}, and active galactic nuclei (AGN) appear to play only a minor role in their formation \citep{Smirnov2020AstL...46..501S}. From the Widefield ASKAP L-band Legacy All-sky Blind surveY (WALLABY; \citealt{Koribalski2020Ap&SS.365..118K}), two PRG candidates with spiral hosts were identified, each containing about 50\% of their total atomic gas in the polar ring \citep{Deg2023MNRAS.525.4663D}.  Based on the observed frequency and model predictions, the incidence rate of PRGs is estimated to be $\sim$1--3\%. Recent MeerKAT data also reveals a potential polar \HI\ disk in NGC~5068 \citep{Healy2024A&A...687A.254H}, and its anomalous gas should be originated from external accretion.

The Tully$-$Fisher relation (TFR) describes the empirical correlation between galaxy rotation velocity and total mass \citep{Tully1977A&A....54..661T}, the latter often represented by absolute magnitude \citep{Giovanelli1997AJ....113...22G, Tully2013AJ....146...86T, Neill2014ApJ...792..129N}, stellar mass \citep{Bell2001ApJ...550..212B}, or baryonic mass \citep{McGaugh2000ApJ...533L..99M, Lelli2019MNRAS.484.3267L}. The tight scatter ($\sim$0.2 dex; \citealt{Bradford2016ApJ...832...11B}) indicates a close link between galactic dynamics and baryonic growth. Because of their extended rings, PRGs provide unique laboratories for probing the kinematics of dark matter halos and measuring rotation in massive early-type systems, which may deviate from the canonical TFR of normal galaxies.

A notable break in the TFR occurs below rotation velocities of $\sim$90~km~s$^{-1}$, where galaxies appear underluminous relative to the high-mass extrapolation \citep{McGaugh2000ApJ...533L..99M}. When the gas mass is included, however, both low- and high-mass galaxies align along a single baryonic TFR (bTFR; \citealt{McGaugh2000ApJ...533L..99M, McGaugh2005ApJ...632..859M}). Moreover, the bTFR slope provides a diagnostic for cosmological models: values near 3 are consistent with $\Lambda$CDM predictions, while slopes near four support modified Newtonian dynamics (MOND; \citealt{McGaugh2012AJ....143...40M}). Thus, examining the bTFR offers constraints on galaxy formation and cosmological evolution.

PRGs and PRG candidates exhibit higher rotation velocities than spiral galaxies of comparable luminosity \citep{Iodice2003ApJ...585..730I, Reshetnikov2004A&A...416..889R}, yet follow a TFR similar to lenticular (S0) systems \citep{Knapp1985A&A...142....1K}. Although spirals and S0s share similar slopes, S0 galaxies are on average 0.5~mag fainter in the $K_S$ band at a given rotation velocity \citep{Williams2010MNRAS.409.1330W}, suggesting a possible evolutionary link. However, the intrinsic scatter and slope of TFRs also depend on rotation velocity definitions, fitting methods, sample selection \citep{Bradford2016ApJ...832...11B}, and tracers (e.g., CO, H$\alpha$, or \HI; \citealt{Lelli2019MNRAS.484.3267L}). Large, homogeneous datasets are therefore essential for disentangling these effects and quantifying systematic differences.

The Five-hundred-meter Aperture Spherical Telescope (FAST; \citealt{Nan2011IJMPD..20..989N, Li2018IMMag..19..112L}) provides an unprecedented opportunity to explore these questions. In this work, we perform uniform \HI\ observations of dozens of kinematically confirmed PRGs and compile additional data from the literature. We then compare the TFRs of PRGs with those of other galaxy types to constrain their formation mechanisms. Section~\ref{sec:sample} presents the PRG sample, FAST data, and control galaxies. Section~\ref{sec:sf-gas} compares the colors and gas fractions of PRGs with those of the control sample. Section~\ref{sec:TFR} examines the TFRs, and Section~\ref{sec:sum} summarizes the results. Throughout this paper, we adopt a flat $\Lambda$CDM cosmology with $\Omega_m = 0.315$, $\Omega_\Lambda = 0.685$, and $H_0 = 67.4$~km~s$^{-1}$~Mpc$^{-1}$ \citep{PlanckCollaboration2020AA...641A...6P}, and assume a Chabrier initial mass function (IMF; \citealt{Chabrier2003PASP..115..763C}) to derive stellar masses and star formation rates uniformly.

\section{PRG Sample \& Data}
\label{sec:sample}

We compile all kinematically confirmed PRGs from the literature. To probe their \HI\ gas dynamics, we conducted new \HI\ observations for 22\footnote{Two of them (PRC~C-24 and SPRC~201) are not included because lack of kinematically evidence for PRG identification.} of them in our follow-up campaign utilizing FAST, and collected the \HI\ measurements results on the rest of the PRGs when available. Combining archival and newly obtained data, we present their \HI\ spectra and apply the curve-of-growth method \citep{Yu2020ApJ...898..102Y, Yu2022ApJS..261...21Y} to measure \HI\ masses, line widths, asymmetry, and profile shape. Our final sample with \HI\ measurements consists of 40 kinematically confirmed PRGs.

\subsection{The sample of Kinematically confirmed PRGs}
\label{subsec:prg}

We have assembled a comprehensive sample of kinematically confirmed PRGs. To date, approximately 400 PRGs and PRG candidates have been cataloged across various studies \citep{Whitmore1987ApJ...314..439W, Whitmore1990AJ....100.1489W, Reshetnikov1995AA...303..398R, Cox1995NYASA.751...27C, Iodice2003ApJ...585..730I, Battinelli2006AA...451...99B, Moiseev2011MNRAS.418..244M, Merkulova2012AstBu..67..374M, Moiseev2014ASPC..486...71M, Reshetnikov2019MNRAS.483.1470R}, mainly from the visual identification based on optical images. But it is fraught with high uncertainty, because the perpendicular structure can be caused by the overlap of two galaxies, such as SPRC~178 \citep{Moiseev2011MNRAS.418..244M}. We collect all kinematically confirmed PRGs following the definition in \citet{Whitmore1990AJ....100.1489W}, which exhibit significant misalignment in galaxy rotation along the major and minor axes, such as SPRC~260 in \citet{Moiseev2014ASPC..486...71M}. This curated sample of kinematically confirmed PRGs is essential for elucidating the intrinsic properties and unraveling the formation mechanisms of these enigmatic galaxies. 

Our sample encompasses 40 PRGs with stellar masses ranging from 8.8 to 11.0 and redshifts spanning from 0 to 0.08. The fundamental properties of these galaxies are documented in Table~\ref{tab:basic}, constituting the most extensive and complete compilation of kinematically confirmed PRGs to date. 

\begin{deluxetable*}{llrrrrrr}
\tablenum{1}
\tabletypesize{\footnotesize}
\tablecolumns{8}
\tablecaption{A comprehensive database of 40 kinematically confirmed PRGs.}
\tablehead{
\colhead{PRG} &
\colhead{Name} &
\colhead{R.A.} &
\colhead{Decl.} &
\colhead{$z$} &
\colhead{References}  &
\colhead{Mode} &
\colhead{Time} 
\\
\colhead{} &
\colhead{} &
\colhead{(J2000)} &
\colhead{(J2000)} &
\colhead{} &
\colhead{} &
\colhead{} &
\colhead{(second)} 
\\
\colhead{(1)} &
\colhead{(2)} &
\colhead{(3)} &
\colhead{(4)} &
\colhead{(5)} &
\colhead{(6)} &
\colhead{(7)} &
\colhead{(8)} 
}
\startdata
PRC~A-01 & PGC~006101 & 24.73014 & -7.76546 & 0.01844 & 1 & Tracking & 3600 \\
PRC~A-02 & ESO~415-G~026 & 37.08376 & -31.88100 & 0.01536 & 1, 2 &    &  \\
PRC~A-03 & NGC~2685 & 133.89470 & 58.73442 & 0.00294 & 1 & MultiBeamOTF & 33408 \\
PRC~A-04 & UGC~07576 & 186.92441 & 28.69806 & 0.02341 & 1 & Tracking & 1800 \\
PRC~A-05 & NGC~4650A & 191.20439 & -40.71432 & 0.00961 & 1, 2 &    &  \\
PRC~A-06 & UGC~9796 & 228.98462 & 43.16679 & 0.01798 & 1 & Tracking & 900 \\
PRC~B-01 & IC~51 & 11.60097 & -13.44224 & 0.00574 & 3 & MultiBeamOTF & 3060 \\
PRC~B-03 & IC~1689 & 20.94950 & 33.05532 & 0.01523 & 4 & Tracking & 1800 \\
PRC~B-09 & UGC~05119 & 144.30273 & 38.09239 & 0.02005 & 5 &    &  \\
PRC~B-16 & NGC~5122 & 201.06225 & -10.65423 & 0.00940 & 6, 7 & Tracking & 900 \\
PRC~B-17 & UGC~9562 & 222.81009 & 35.54232 & 0.00426 & 8 & Tracking & 900 \\
PRC~B-18 & PGC~089058 & 294.66002 & -56.45796 & 0.03886 & 9 &    &  \\
PRC~B-19 & AM~2020-504 & 305.97878 & -50.65187 & 0.01695 & 10 &    &  \\
PRC~B-21 & ESO~603-G21 & 342.84188 & -20.24734 & 0.01042 & 11, 12 &    &  \\
PRC~C-03 & ESO~474-G26 & 11.78138 & -24.37074 & 0.05271 & 13 &    &  \\
PRC~C-13 & NGC~660 & 25.76004 & 13.64504 & 0.00283 & 14 &    &  \\
PRC~C-27 & UGC~4385 & 125.96664 & 14.75185 & 0.00657 & 8 & Tracking & 900 \\
PRC~C-28 & NGC~2748 & 138.42916 & 76.47538 & 0.00492 & 8, 15 &    &  \\
PRC~C-45 & NGC~5128 & 201.36506 & -43.01911 & 0.00182 & 16 &    &  \\
PRC~C-46 & ESO~576-~G~069 & 202.52204 & -20.93306 & 0.01781 & 7 &    &  \\
SPRC~001 & MaNGA~01-287487 & 2.29825 & -0.61520 & 0.07333 & 17 &    &  \\
SPRC~002 & PGC~3089465 & 6.67298 & 25.85048 & 0.02850 & 17 &    &  \\
SPRC~003 & NSA~022108 & 10.26435 & -9.94108 & 0.03683 & 17, 18 & Tracking & 5100 \\
SPRC~004 & NSA~007324 & 40.74342 & -0.95259 & 0.04301 & 18 &    &  \\
SPRC~007 & LEDA~3444084 & 118.14304 & 29.34713 & 0.06007 & 19 &    &  \\
SPRC~010 & MaNGA~01-460660 & 125.15907 & 15.61664 & 0.04244 & 20 & Tracking & 5400 \\
SPRC~012 & NaN & 135.45893 & 52.20626 & 0.06247 & 17 &    &  \\
SPRC~013 & MaNGA~01-124268 & 138.72352 & 49.63998 & 0.03177 & 17 &    &  \\
SPRC~014 & CGCG~121-053 & 139.56654 & 20.36816 & 0.03188 & 20 & Tracking & 1800 \\
SPRC~027 & NSA~113235 & 176.18349 & 23.16248 & 0.04835 & 17, 18 & Tracking & 4200 \\
SPRC~033 & NGC~4262 & 184.87741 & 14.87765 & 0.00453 & 21 & Tracking & 900 \\
SPRC~037 & LEDA~2354098 & 193.29069 & 49.80903 & 0.06762 & 17 &    &  \\
SPRC~039 & NSA~056186 & 197.07056 & 45.37643 & 0.02933 & 20 & Tracking & 3600 \\
SPRC~040 & NGC~5014 & 197.87994 & 36.28200 & 0.00379 & 17 & Tracking & 900 \\
SPRC~060 & ASK~404109.0 & 236.85131 & 38.93064 & 0.07828 & 20 &    &  \\
SPRC~067 & CGCG~225-097 & 259.43391 & 40.69779 & 0.02777 & 22 & Tracking & 900 \\
SPRC~069 & II~Zw~092 & 312.02359 & 0.06860 & 0.02469 & 20 & Tracking & 1800 \\
SPRC~241 & MCG~-01-07-035 & 40.48751 & -6.79226 & 0.01757 & 18 & Tracking & 1800 \\
SPRC~260 & CGCG~068-056 & 176.37606 & 9.72909 & 0.02135 & 18 & MultiBeamOTF & 4560 \\
NGC~7625 & Arp~212 & 350.12535 & 17.22565 & 0.00543 & 23 &    &  \\
\enddata
\tablecomments{
Column (1): the galaxy name from the PRG catalogs. Column (2): a common galaxy name. Columns (3)--(4): equatorial coordinates (J2000) from NASA/IPAC Extragalactic Database (NED). Column (5): optical redshift from NED with a typical uncertainty of a few percent. Column (6): references for the kinematical confirmation.  
1: \citet{Whitmore1990AJ....100.1489W},  
2: \citet{Whitmore1987ApJ...314..439W},  
3: \citet{Schiminovich2013AJ....145...34S},  
4: \citet{Hagen-Thorn1997AA...319..430H},  
5: \citet{Merkulova2008AstL...34..542M},  
6: \citet{Cox1995NYASA.751...27C},  
7: \citet{Reshetnikov2001MNRAS.322..689R},  
8: \citet{Reshetnikov1994AA...291...57R},  
9: \citet{Reshetnikov2006AA...446..447R},  
10: \citet{Arnaboldi1993AA...267...21A},  
11: \citet{Arnaboldi1994ASPC...54..437A},  
12: \citet{Reshetnikov2002AA...383..390R},  
13: \citet{Reshetnikov2005AA...431..503R},  
14: \citet{vanDriel1995AJ....109..942V},  
15: \citet{Merkulova2009AstL...35..587M},  
16: \citet{Schiminovich1994ApJ...423L.101S},  
17: \citet{Egorov2019MNRAS.486.4186E}, 
18. \citet{Moiseev2014ASPC..486...71M}, 
19: \citet{Brosch2010MNRAS.401.2067B}, 
20: \citet{Moiseev2011MNRAS.418..244M}, 
21: \citet{Bettoni2010AA...519A..72B},  
22: \citet{Merkulova2013arXiv1302.1339M},  
23. \citet{Moiseev2008AstBu..63..201M}. 
Column (7): the FAST observation mode. Column (8): the FAST integration time in unit of second.
}
\label{tab:basic}
\end{deluxetable*}

\begin{figure*}[h]
\centering
\epsscale{1.1}
\plotone{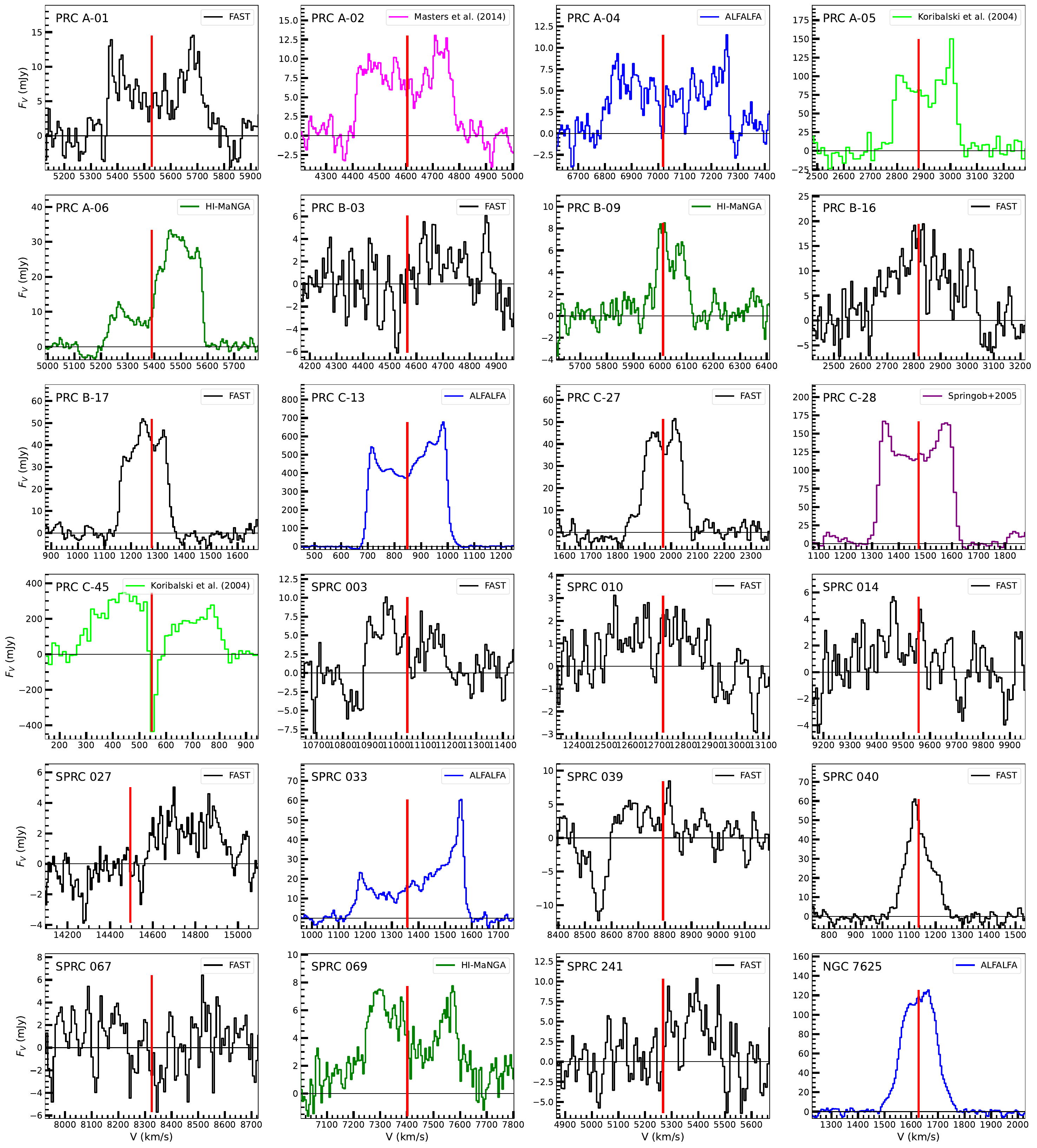}
\caption{
The \HI\ spectra of PRGs. In each panel, the galaxy name is shown in the upper left corner, and the optical central velocity is shown as the red vertical line. The colors distinguish the origin of \HI\ spectra: black (FAST or this work), magenta \citep{Masters2014MNRAS.443.1044M}, blue \citep{Haynes2018ApJ...861...49H}, lime \citep{Koribalski2004AJ....128...16K}, green \citep{Stark2021MNRAS.503.1345S}, and violet \citep{Springob2005ApJS..160..149S}.
}
\label{fig:fast-sp}
\end{figure*}

\subsection{FAST \HI\ Observations}
\label{subsec:fast}

We obtained \HI\ spectral line observations for 20 kinematically confirmed PRGs using FAST. The sample selection was constrained by FAST's observable sky coverage and the requirement that the integration time for each target not exceed one hour, except for NGC~2685 which is included for our interest in its detailed analysis. We observed our PRG targets during FAST's ``share-risk'' observation period in 2019 (PI: Niankun Yu, Project code: 2019a-047-S) utilizing the Onoff mode. However, due to significant Radio Frequency Interference (RFI) contamination, only two reliable detections were obtained. Consequently, we observe the rest of the sample in 2022 (PI: Pei Zuo, project code: PT2022\_0099; PI: Niankun Yu, project code: PT2022\_0147).

Our FAST observations used employed the 19-beam receiver with the tracking and Multi-beam OTF modes, depending on the angular sizes of our targeted PRGs. The \HI\ radius for each galaxy was estimated using the \HI\ mass-size relation \citep{Broeils1997A&A...324..877B, Wang2016MNRAS.460.2143W}. Considering FAST beam angular size of 2.9$^{\prime}$ at 1.4 GHz, the multi-beam OTF mode were used for three galaxies (CGCG~068-056, IC~51, and NGC~2685) with \HI\ radius larger than the FAST beam size. We map the \HI\ within these three galaxies and their local environment. 
The remaining galaxies were observed in the tracking mode, with beam M14 (offset by 11.6$^{\prime}$ from the center of the FAST's central beam M01) used as the off-point during data processing.

The observations spanned the entire wide bandwidth of 500~MHz at the L~band (1.05--1.45~GHz) with a raw spectral channel width of 7.6~kHz ($\sim$1.6~km~s$^{-1}$). For tracking observations, the sampling time is set to 1.0 second with a 10 K (high CAL) noise diode activated for the first 5 seconds of each cycle and deactivated for the remaining 300 seconds. The Multi-beam OTF observations featured a sampling time of 0.5 seconds, with the noise diode on for the first 1 second and off for the subsequent 99 seconds, scanning the sky at a rate of 5$^{\prime\prime}$ per second. The scanning time is 4560 seconds for CGCG~068-056 and 3060 seconds for IC~51, respectively.

\begin{figure*}
\epsscale{1.}
\plotone{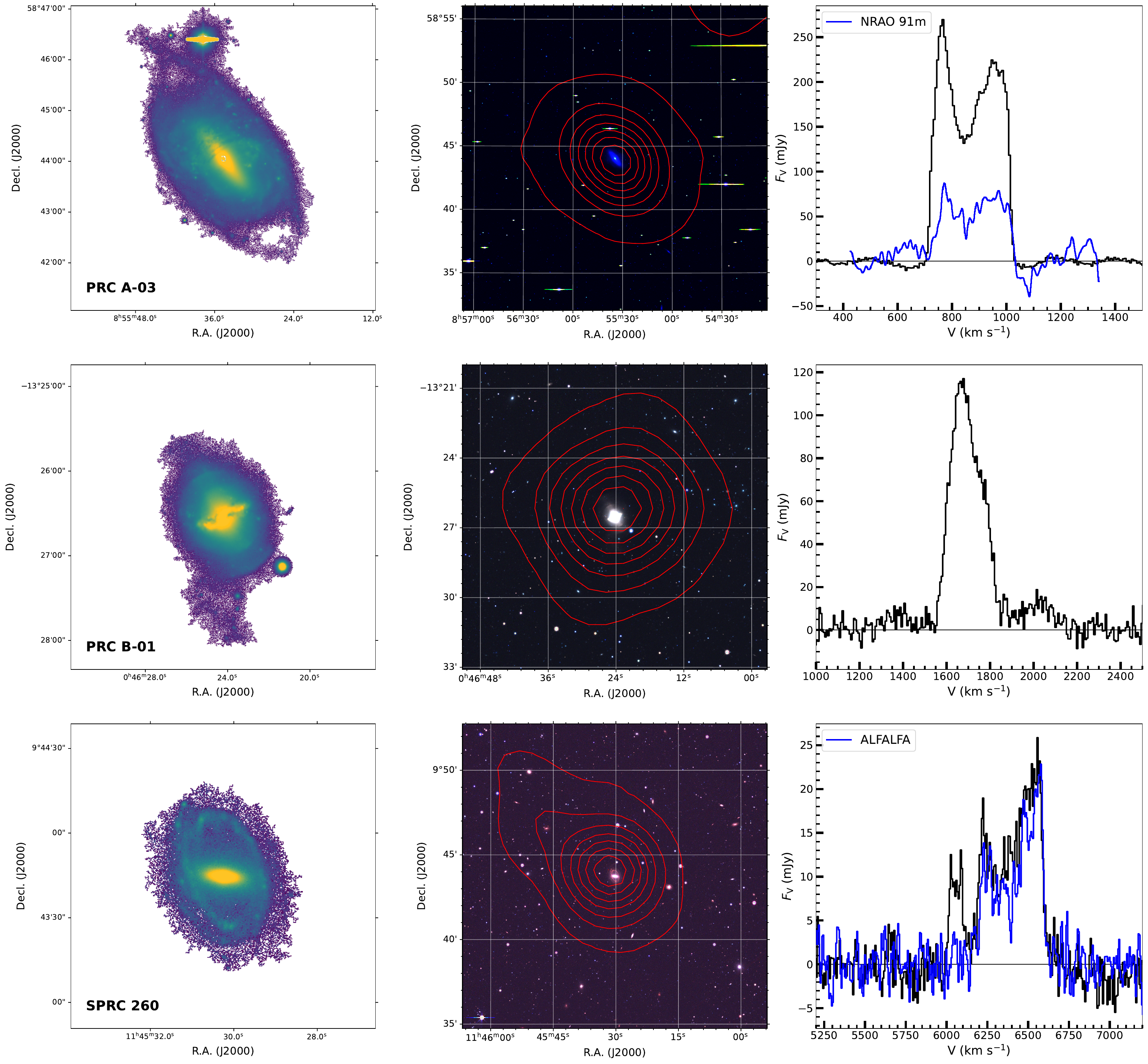}
\caption{
The optical image and \HI\ contours/profiles for three mapping galaxies (Top: PRC~A-03, middle: PRC~B-01, bottom: SPRC~260
).
In each row, the left panel shows the $g$-band image from the DESI survey \citep{Dey2019AJ....157..168D} following the method in \citet{Akhlaghi2015ApJS..220....1A} to highlight the galaxy structure, the middle panel represents the zoom-out optical composite image overlaid with the \HI\ contours (the outer one is the 2$\sigma$ level), and the right panel displays the integrated \HI\ spectrum from FAST (black). In the top and bottom right panel, the blue spectrum is from \citet{Peterson1974AJ.....79..767P} and the ALFALFA survey \citep{Haynes2018ApJ...861...49H}, respectively.
}
\label{fig:fast-mapping}
\end{figure*}

In summary, a total of 20 galaxies were observed with the FAST 19-beam L-band receiver: three in Multi-beam OTF mode and 17 in tracking mode. 
Of the 20 FAST-observed galaxies, 12 show emission-line detections, one exhibits a potential absorption feature (SPRC~039), and 7 are non-detections. The resulting \HI\ spectra are displayed as red profiles in Figure~\ref{fig:fast-sp} and in the right panels of Figure~\ref{fig:fast-mapping}.
For galaxies with multiple observations, we select the spectrum with the optimal beam size and highest S/N. The full dataset thus comprises 27 \HI\ information: 22 detections, 1 absorption, and 4 non-detections (Table~\ref{tab:co-add}).

\subsection{FAST Data Processing}
\label{sec:reduction}

Data processing for the tracking observations followed the procedures described in \citet{Yu2024SCPMA..6799811Y}, with improvements in RFI flagging and baseline subtraction. A frequency window of $\pm10$~MHz around each galaxy's central frequency was extracted, and the flux intensity was calibrated using the on$-$off noise diode data. Aperture efficiency corrections were applied to account for beam, frequency, and positional variations. To optimize observing time, we used other beams of the 19-beam receiver (e.g., M14) as off-target references. We manually flagged and removed RFI-contaminated rows from the raw data. For the remaining data, we subtracted baselines and standing waves for each one-second integration. The XX and YY polarization spectra were then averaged to produce the preliminary \HI\ spectrum for each observation. For sources with multiple observations, we combined the spectra using inverse noise weighting \citep{Fabello2011MNRAS.411..993F, Brown2015MNRAS.452.2479B}. The final spectra obtained in the tracking mode are shown in Figure~\ref{fig:fast-sp}.

The Multi-beam OTF data were processed using a similar workflow to that of the tracking mode (Yu~et~al., in prep.), with additional manual RFI and signal masking. The data were calibrated using the on$-$off noise diode data, followed by median filtering. After RFI masking and baseline subtraction, the spectra from all 19 beams were regrided for both R.A.\ and Decl.\ scans with \textcolor{purple}{$45^{\prime}\times 40^{\prime}$} spaxel to construct \HI\ data cubes representing the spatial distribution of the emission. The resulting intensity maps for the two galaxies observed in OTF mode are shown in Figure~\ref{fig:fast-mapping}. We also present the \HI\ distribution and global spectrum of NGC~2685 in the same figure. All three mapping observations reveal significant \HI\ emission; both CGCG~068--056 and NGC~2685 show additional detections in their companion galaxies. All four companions of NGC~2685 are detected in \HI, and CGCG~068--056 has one companion detected with a centroid velocity of $\sim$6100~\kms.

As illustrated in Figure~\ref{fig:fast-sp}, the FAST spectra generally show lower flux intensities than those reported in the literature, which are typically derived from observations with larger integration areas (no smaller than 7$^{\prime}\times$7$^{\prime}$, \citealt{Haynes2018ApJ...861...49H}). The FAST tracking observations, with a 2.9$^{\prime}$ beam, may therefore miss extended \HI\ emission. In contrast, the Multi-beam OTF observations fully cover the extended regions and yield fluxes consistent with previous measurements (Figure~\ref{fig:fast-mapping}), confirming that the apparent flux deficit in the tracking spectra arises from the smaller FAST beam.

\subsection{Archival \HI\ Spectra}
In addition to the \HI\ data from our FAST observations, we compiled PRG \HI\ spectra from the literature \citep{Haynes2018ApJ...861...49H, Stark2021MNRAS.503.1345S, Springob2005ApJS..160..149S, Koribalski2004AJ....128...16K, Masters2014MNRAS.443.1044M}. All spectra are shown in Figures~\ref{fig:fast-sp} and \ref{fig:fast-mapping}. Both the new FAST and archival spectra were RFI-masked, baseline-subtracted, and rebinned to a channel width of $\sim$20 km s$^{-1}$. A detection in this work is defined as at least three consecutive channels with flux densities exceeding 3$\sigma_{\rm spec}$, where $\sigma_{\rm spec}$ is derived from a half-Gaussian fit to distribution of the negative flux values within $\pm$500 km s$^{-1}$ to the systemic velocity from optical spectroscopy.

We also incorporate \HI\ total fluxes for six PRGs and the spectral noise for one PRG from prior surveys \citep{Richter1994AJ....107...99R, vanDriel2002AA...386..140V}. Lacking full spectra, these data are used only to estimate \HI\ masses or upper limits. A complete summary of \HI\ measurements and derived quantities is given in Table~\ref{tab:co-add}.

\begin{deluxetable*}{lrrrrrrrrrrrrrrrr}
\tablenum{2}
\rotate
\tiny
\tablecaption{Properties of Polar Ring Galaxies}
\label{tab:co-add}
\tablehead{
\colhead{Galaxy} &
\colhead{$D_{\rm L}$} &
\colhead{$V_c$} &
\colhead{$F$} &
\colhead{$V_{85}$} &
\colhead{$A_F$} &
\colhead{$K$} &
\colhead{S/N} &
\colhead{$\sigma$} &
\colhead{log~$M_{\rm HI}$} &
\colhead{$i$} &
\colhead{$V_{\rm rot}$} &
\colhead{$V_{\rm rot,\,i=90}$} &
\colhead{Ref} &
\colhead{NUV$-$r} &
\colhead{log~$M_\star$} &
\colhead{log~$M_{\rm bary}$} \\
& (Mpc) & (km s$^{-1}$) & (Jy km s$^{-1}$) & (km s$^{-1}$) &  &  &  &
(mJy) & ($M_\odot$) & (deg) & (km s$^{-1}$) & (km s$^{-1}$) &  &
(mag) & ($M_\odot$) & ($M_\odot$)
}
\startdata
PRC~A-01 & 83 & 5557 $\pm$ 8 & 2.70 $\pm$ 0.42 & 336 $\pm$ 21 & 1.03 $\pm$ 0.08 & -0.072 $\pm$ 0.019 & 14.5 & 2.4 & 9.6 $\pm$ 0.1 & 40 $\pm$ 7 & 259 $\pm$ 38 & 171 $\pm$ 10 & 1 & 4.3 & 10.1 & 10.3 $\pm$ 0.3 \\
PRC~A-02 & 69 & 4601 $\pm$ 5 & 3.05 $\pm$ 0.48 & 320 $\pm$ 19 & 1.03 $\pm$ 0.07 & -0.049 $\pm$ 0.016 & 30.0 & 1.5 & 9.5 $\pm$ 0.1 &    &    & 164 $\pm$ 9 & 2 &  &  &  \\
PRC~A-03 & 13 & 867 $\pm$ 3 & 54.85 $\pm$ 8.67 & 245 $\pm$ 15 & 1.06 $\pm$ 0.07 & -0.049 $\pm$ 0.015 & 194.3 & 4.4 & 9.3 $\pm$ 0.1 & 60 $\pm$ 6 & 147 $\pm$ 12 & 128 $\pm$ 7 & 1 & 4.4 & 10.2 & 10.3 $\pm$ 0.3 \\
PRC~A-04 & 106 & 7034 $\pm$ 6 & 2.47 $\pm$ 0.39 & 426 $\pm$ 28 & 1.02 $\pm$ 0.08 & -0.063 $\pm$ 0.018 & 24.2 & 1.4 & 9.8 $\pm$ 0.1 & 48 $\pm$ 7 & 284 $\pm$ 34 & 214 $\pm$ 13 & 3 & 4.8 & 10.4 & 10.5 $\pm$ 0.3 \\
PRC~A-05 & 43 & 2909 $\pm$ 4 & 22.53 $\pm$ 3.55 & 217 $\pm$ 13 & 1.03 $\pm$ 0.08 & -0.048 $\pm$ 0.016 & 27.6 & 8.7 & 10.0 $\pm$ 0.1 &    &    & 115 $\pm$ 6 & 4 &  &  &  \\
PRC~A-06 & 81 & 5451 $\pm$ 3 & 6.36 $\pm$ 1.01 & 255 $\pm$ 16 & 1.56 $\pm$ 0.11 & 0.066 $\pm$ 0.016 & 67.8 & 1.5 & 10.0 $\pm$ 0.1 & 53 $\pm$ 6 & 164 $\pm$ 16 & 133 $\pm$ 7 & 5 & 4.5 & 10.1 & 10.4 $\pm$ 0.3 \\
PRC~B-01 & 26 & 1728 $\pm$ 3 & 22.62 $\pm$ 3.58 & 284 $\pm$ 19 & 1.66 $\pm$ 0.12 & 0.076 $\pm$ 0.018 & 79.7 & 3.6 & 9.5 $\pm$ 0.1 &    &    & 147 $\pm$ 9 & 1 &  &  &  \\
PRC~B-03 & 68 &    &    &    &    &    &  & 2.3 & $\leq$8.4 & 53 $\pm$ 6 &    &    & 1 & 4.7 & 10.4 &  \\
PRC~B-09 & 90 & 6035 $\pm$ 4 & 0.82 $\pm$ 0.13 & 119 $\pm$ 9 & 1.12 $\pm$ 0.09 & -0.002 $\pm$ 0.018 & 15.9 & 1.2 & 9.2 $\pm$ 0.1 & 48 $\pm$ 7 & 89 $\pm$ 10 & 69 $\pm$ 4 & 5 & 5.2 & 10.6 & 10.7 $\pm$ 0.3 \\
PRC~B-16 & 42 & 2840 $\pm$ 9 & 3.67 $\pm$ 0.57 & 321 $\pm$ 22 & 1.45 $\pm$ 0.12 & 0.065 $\pm$ 0.018 & 13.3 & 3.7 & 9.2 $\pm$ 0.1 &    &    & 164 $\pm$ 10 & 1 &  &  &  \\
PRC~B-17 & 19 & 1254 $\pm$ 3 & 7.47 $\pm$ 1.18 & 151 $\pm$ 9 & 1.02 $\pm$ 0.07 & 0.024 $\pm$ 0.015 & 53.6 & 2.6 & 8.8 $\pm$ 0.1 & 36 $\pm$ 7 & 135 $\pm$ 23 & 84 $\pm$ 4 & 1 & 2.5 & 8.9 & 9.2 $\pm$ 0.3 \\
PRC~C-13 & 13 & 859 $\pm$ 3 & 148.21 $\pm$ 23.43 & 267 $\pm$ 16 & 1.12 $\pm$ 0.08 & -0.029 $\pm$ 0.015 & 543.3 & 4.4 & 9.7 $\pm$ 0.1 & 67 $\pm$ 6 & 150 $\pm$ 10 & 139 $\pm$ 8 & 3 & 4.5 & 10.0 & 10.2 $\pm$ 0.3 \\
PRC~C-27 & 29 & 1967 $\pm$ 3 & 6.90 $\pm$ 1.09 & 140 $\pm$ 9 & 1.08 $\pm$ 0.08 & -0.009 $\pm$ 0.016 & 42.1 & 2.9 & 9.1 $\pm$ 0.1 & 44 $\pm$ 7 & 107 $\pm$ 13 & 79 $\pm$ 4 & 1 & 1.8 & 9.3 & 9.6 $\pm$ 0.3 \\
PRC~C-28 & 22 & 1467 $\pm$ 3 & 39.30 $\pm$ 6.21 & 251 $\pm$ 15 & 1.02 $\pm$ 0.07 & -0.036 $\pm$ 0.015 & 143.3 & 3.1 & 9.6 $\pm$ 0.1 &    &    & 131 $\pm$ 7 & 8 &  &  &  \\
PRC~C-45 & 8 & 527 $\pm$ 17 & 97.53 $\pm$ 16.31 & 484 $\pm$ 34 & 2.00 $\pm$ 0.29 & -0.046 $\pm$ 0.049 & 17.3 & 52.0 & 9.2 $\pm$ 0.1 &    &    & 241 $\pm$ 16 & 4 &  &  &  \\
SPRC~001 & 343 &    &    &    &    &    &  &  &    & 48 $\pm$ 7 &    &    &  & 5.1 & 11.0 &    \\
SPRC~003 & 168 & 11019 $\pm$ 17 & 1.25 $\pm$ 0.23 & 230 $\pm$ 33 & 1.13 $\pm$ 0.30 & 0.013 $\pm$ 0.044 & 6.5 & 2.7 & 9.9 $\pm$ 0.1 & 19 $\pm$ 11 & 351 $\pm$ 193 & 121 $\pm$ 15 & 1 & 3.7 & 10.0 & 10.3 $\pm$ 0.3 \\
SPRC~004 & 197 &    &    &    &    &    &  &  &    & 47 $\pm$ 7 &    &    &  & 3.5 & 9.8 &    \\
SPRC~007 & 278 &    &    &    &    &    &  &  &    & 45 $\pm$ 7 &    &    &  & 3.2 & 10.2 &    \\
SPRC~010 & 194 &    &    &    &    &    &  & 1.0 & $\leq$9.0 & 58 $\pm$ 7 &    &    & 1 & 5.1 & 10.3 &  \\
SPRC~013 & 144 &    &    &    &    &    &  &  &    & 43 $\pm$ 7 &    &    &  & 3.8 & 10.3 &    \\
SPRC~014 & 145 &    &    &    &    &    &  & 1.7 & $\leq$9.0 & 22 $\pm$ 10 &    &    & 1 & 4.4 & 10.6 &  \\
SPRC~027 & 222 & 14769 $\pm$ 13 & 0.88 $\pm$ 0.16 & 306 $\pm$ 37 & 1.00 $\pm$ 0.14 & -0.027 $\pm$ 0.032 & 9.0 & 1.4 & 10.0 $\pm$ 0.1 & 28 $\pm$ 8 & 324 $\pm$ 95 & 157 $\pm$ 18 & 1 & 3.7 & 10.3 & 10.5 $\pm$ 0.3 \\
SPRC~033 & 20 & 1413 $\pm$ 4 & 8.55 $\pm$ 1.35 & 346 $\pm$ 21 & 1.34 $\pm$ 0.09 & -0.023 $\pm$ 0.018 & 74.1 & 1.7 & 8.9 $\pm$ 0.1 & 28 $\pm$ 8 & 362 $\pm$ 97 & 176 $\pm$ 10 & 3 & 5.5 & 10.3 & 10.3 $\pm$ 0.3 \\
SPRC~037 & 315 &    &    &    &    &    &  &  &    & 20 $\pm$ 11 &    &    &  & 3.3 & 10.6 &    \\
SPRC~039 & 133 &    &    &    &    &    &  & 1.9 &    & 52 $\pm$ 6 &    &    & 1 & 5.6 & 9.9 &  \\
SPRC~040 & 17 & 1137 $\pm$ 3 & 5.85 $\pm$ 0.93 & 132 $\pm$ 8 & 1.17 $\pm$ 0.09 & 0.079 $\pm$ 0.015 & 45.1 & 2.4 & 8.6 $\pm$ 0.1 &    &    & 75 $\pm$ 4 & 1 &  &  &  \\
SPRC~060 & 367 &    &    &    &    &    &  &  &    & 45 $\pm$ 7 &    &    &  & 2.4 & 10.2 &    \\
SPRC~067 & 126 &    &    &    &    &    &  & 2.4 & $\leq$9.0 & 57 $\pm$ 7 &    &    & 1 & 5.3 & 10.6 &  \\
SPRC~069 & 111 & 7398 $\pm$ 4 & 2.22 $\pm$ 0.35 & 427 $\pm$ 28 & 1.15 $\pm$ 0.08 & 0.014 $\pm$ 0.017 & 38.2 & 0.8 & 9.8 $\pm$ 0.1 & 10 $\pm$ 18 & 1127 $\pm$ 1871 & 214 $\pm$ 13 & 5 & 4.4 & 10.2 & 10.4 $\pm$ 0.3 \\
SPRC~241 & 79 & 5382 $\pm$ 13 & 1.01 $\pm$ 0.19 & 196 $\pm$ 31 & 1.00 $\pm$ 0.39 & 0.099 $\pm$ 0.044 & 6.0 & 2.9 & 9.2 $\pm$ 0.1 & 54 $\pm$ 6 & 127 $\pm$ 20 & 105 $\pm$ 15 & 1 & 4.7 & 10.3 & 10.3 $\pm$ 0.3 \\
SPRC~260 & 96 & 6382 $\pm$ 5 & 6.29 $\pm$ 1.00 & 370 $\pm$ 22 & 1.46 $\pm$ 0.11 & -0.046 $\pm$ 0.015 & 43.1 & 1.9 & 10.1 $\pm$ 0.1 & 48 $\pm$ 7 & 246 $\pm$ 28 & 187 $\pm$ 10 & 1 & 3.7 & 10.3 & 10.6 $\pm$ 0.3 \\
NGC~7625 & 24 & 1630 $\pm$ 3 & 19.08 $\pm$ 3.02 & 156 $\pm$ 9 & 1.03 $\pm$ 0.07 & 0.041 $\pm$ 0.015 & 131.6 & 2.7 & 9.4 $\pm$ 0.1 & 35 $\pm$ 7 & 142 $\pm$ 25 & 87 $\pm$ 4 & 3 & 3.5 & 10.1 & 10.2 $\pm$ 0.3 \\
\enddata
\tablecomments{
Column (1): the galaxy name. Column (2): the luminosity distance by adopting cosmology of the Planck Collaboration \citep{PlanckCollaboration2020AA...641A...6P}. Column (3): \HI\ Flux intensity-weighted central velocity. Column (4): Total global flux of the \HI\ line. Column (5): Velocity width measured at 85\% of the total line flux. Column (6): \HI\ profile asymmetry. Column (7): \HI\ profile shape. Column (8): S/N of the profile. Column (9): Noise level of the profile at a channel width of $\sim$6.4 km\,  s$^{-1}$. Column (10): \HI\ mass and its uncertainty or \HI\ mass upper limits by assuming a 3$\sigma$ detection with a line width of 200 km\,  s$^{-1}$. For a distance uncertainty of 10\% and a flux uncertainty of 15\%, the typical uncertainty of log~\MHI\ is 0.1 dex. Column (11): the optical inclination angle in unit of degree. Column (12): the galaxy rotation velocity by assuming optical inclination angle. Column (13): the galaxy rotation velocity by assuming $i$=90$^{\circ}$. Column (14): reference of \HI\ data -- 1: this work; 2: \citep{Masters2014MNRAS.443.1044M}, 3: \citep{Haynes2018ApJ...861...49H}, 4: \citep{Koribalski2004AJ....128...16K}, 5: \citep{Stark2021MNRAS.503.1345S}, 6: \citep{Richter1994AJ....107...99R}, 7: \citep{vanDriel2002AA...386..140V}, and 8: \citep{Springob2005ApJS..160..149S}. Column (15): NUV$-$r color from the NSA catalog. Column (16): the stellar mass from the NSA catalog, and we assume an uncertainty of 0.3~dex because PRGs are peculiar in its optical morphology color. Column (17): the baryonic mass by considering the uncertainties in both \HI\ mass and stellar mass.
}
\end{deluxetable*}

\subsection{The Control Sample}
\label{subsec:control}
We construct control samples for $\sim$ 7,  800 galaxies of different morphology, including ellipticals, S0s, ring galaxies, and spirals from SDSS. These galaxies are cross-matched between the ALFALFA survey \citep{Haynes2011AJ....142..170H, Haynes2018ApJ...861...49H} and the NASA Sloan Atlas (NSA) catalog \citep{Blanton2011AJ....142...31B} using TOPCAT, requiring a maximum positional offset of $3.0^{\prime\prime}$ and a maximum recession velocity difference of 500~km~s$^{-1}$. The global stellar masses from the NSA catalog are derived from spectral energy distribution (SED) fitting using the $kcorrect$ package \citep{Blanton2007AJ....133..734B}, assuming the BC03 single stellar population model \citep{Bruzual2003MNRAS.344.1000B} and a \citet{Chabrier2003PASP..115..763C} initial mass function. The baryonic mass is computed as the sum of the stellar and gas components:
\begin{equation}
\log M_{\rm bary} = \log (M_\star + 1.33 \times M_{\rm HI}),  
\end{equation}
where the factor of 1.33 accounts for the contribution of Helium (such as \citealt{Salucci1999MNRAS.309..923S, Bell2003ApJ...585L.117B}).

We adopt \HI\ measurements based on the curve-of-growth method \citep{Yu2022ApJS..261...21Y}, excluding galaxies without optical counterparts or with \HI\ spectra contaminated by companions. This selection yields a sample of 19,  419 galaxies, hereafter the ALFALFA$-$NSA sample, spanning a stellar mass range of $\sim10^{7.0}$--$10^{11.5}$~M$_\odot$ and redshift $z=0-$0.06. Inclination angles are computed following Section~\ref{subsec:dect}.

We excluded galaxies with inclination angles less than $30^\circ$ because (1) the observed line widths of face-on galaxies do not reliably trace rotation velocities due to dispersion- and turbulence-dominated line broadening, and (2) the de-projected rotation velocity carries large uncertainties, introducing scatter in the Tully--Fisher relation (TFR). We further require \HI\ spectra with S/N $>7$ to minimize uncertainties in measured line widths. After applying these cuts, 13,  956 galaxies remain. This sample is used to investigate secondary parameters contributing to the scatter of the stellar TFR. We examine correlations with optical radius, stellar mass surface density, deviations of gas fraction and star formation rate relative to stellar mass \citep{Catinella2018MNRAS.476..875C}, \HI\ profile asymmetry, and \HI\ profile shape. 

To ensure that the \HI\ line width robustly traces rotation, we select galaxies exhibiting double-horned or flat-topped profiles, following the criterion for profile shape $K \le 0.02$ established by \citet{Yu2022ApJS..261...21Y}. To reduce the effects of gas deficiency and turbulence, we further require that galaxies have a gas fraction deviation of less than 0.4~dex, based on the green dashed line in Figure~\ref{fig:gf}. Applying these criteria results in a final control sample of 7,  817 galaxies. We do not impose additional cuts on stellar mass or redshift because the control sample already spans a similar mass range and the TFR shows no significant evolution for $z<0.1$ \citep{Portinari2007MNRAS.375..913P}.

\section{Results}
\label{sec:sf-gas}
\subsection{Detections}
\label{subsec:dect}
We measure the total flux and line width of \HI\ emission-line spectra using the curve-of-growth method \citep{Yu2020ApJ...898..102Y, Yu2022ApJS..261...21Y}. For each galaxy, we search for \HI\ emission within $\pm$500 km~s$^{-1}$ of the optical systemic velocity. After identifying channels with potential emission, we construct the curve of growth by integrating flux from the flux intensity-weighted center ($V_{\rm c}$) outward to both sides of the spectrum. The total flux, $F$ (in Jy~km~s$^{-1}$), is taken as the median integrated flux of the flat portion of the curve, and the line width, $V_{85}$, is defined as the velocity interval enclosing 85\% of the total flux. The \HI\ profile asymmetry $A_F \geq 1$ is defined as the ratio of the integrated fluxes of the blue-shift and red-shift sides. The \HI\ profile shape $K$ is defined as the integrated area between the normalized curve of growth and the diagonal line after normalizing the velocity axis by $V_{85}$ and the integrated flux axis by 85\% of the total flux. The results are summarized in Table~\ref{tab:co-add}.

\begin{figure*}[!hbt]
\epsscale{1.05}
\plotone{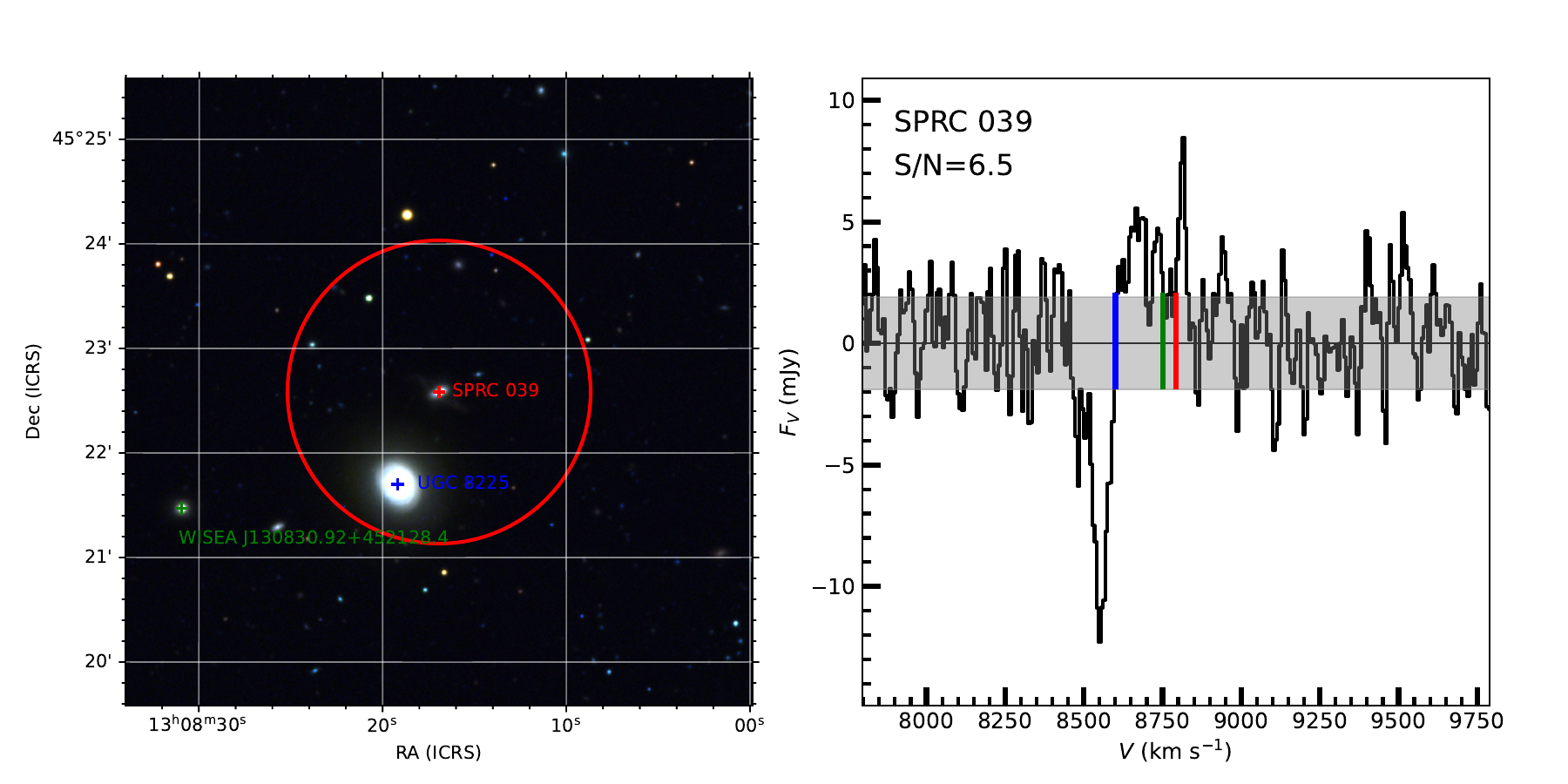}
\caption{Left: the optical composite image of SPRC~39 and its two companions with the red circle show the FAST beam size. Right: The FAST \HI\ spectrum of this pointing, and we mark the central velocities of these three galaxies as vertical lines.
}
\label{fig:absorption}
\end{figure*}

The signal-to-noise ratio (S/N) of each spectrum is calculated as
\begin{equation}
S/N = \frac{1000 F}{\sigma \Delta V \sqrt{2N}},  
\end{equation}
where $2N$ is the number of channels used in the curve-of-growth integration, $\Delta V$ is the channel width in km\,  s$^{-1}$, and $\sigma$ is the noise level in mJy. For spectra with $S/N < 40$, we apply a second-order polynomial correction to account for systematic deviation in the total flux and line width following \citet{Yu2022ApJS..261...21Y}. All reported \HI\ parameters have been corrected for this bias. The \HI\ mass is then calculated as
\begin{equation}
M_{\rm HI} = 2.356\times10^5 D_L^2 F\ M_\odot,  
\end{equation}
where $D_{L}$ is luminosity distance in Mpc \citep{Roberts1962}. The reported line widths are corrected for redshift, instrumental broadening, turbulent motions, and projection effects.

Deprojection of the \HI\ line width assumes that (1) the stellar and \HI\ disks are coplanar and (2) the intrinsic axis ratio $q_0$ of the optical disk depends on stellar mass \citep{Sanchez-Janssen2010MNRAS.406L..65S} and redshift \citep{Chang2013ApJ...773..149C}. The inclination angle is estimated following \citet{Hubble1926ApJ....64..321H}:
\begin{equation}
\cos^2 i = \frac{q^2 - q_0^2}{1 - q_0^2},  
\end{equation}
where $q = b/a$ is the observed axis ratio. We adopt the      $r$-band axis ratio ``PETRO\_BA90'' and Petrosian radius ``PETRO\_THETA'' from the NSA catalog \citep{Blanton2011AJ....142...31B}. The intrinsic axis ratio $q_0$ is estimated for each redshift bin by fitting a second-order polynomial to the 1\% lower envelope of the $q-$stellar mass distribution. The polynomial coefficients are provided in Table~\ref{tab:ba}. The characteristic upturn of $q_0$ at $\log(M_\star/M_\odot) \gtrsim 9.5$ is consistent across all redshift bins, reflecting the ``U-shaped'' trend reported by \citet{Sanchez-Janssen2010MNRAS.406L..65S}, and indicates that massive galaxies generally have thicker disks regardless of redshift.

We estimate rotation velocities for both the control sample and PRGs using the corrected \HI\ line width ($V_{85c}$) and the relation with the rotation curve \citep{Yu2020ApJ...898..102Y}:
\begin{equation}
V_{\rm rot} = (0.94 \pm 0.02) V_{85c} + (13.33 \pm 3.31), 
\end{equation}
with an intrinsic scatter of 27~\kms. PRGs in our sample have \HI\ masses of $10^{8.8}-10^{11.0}$~M$_\odot$ and rotation velocities ranging from 60~km~s$^{-1}$ to 370~km~s$^{-1}$.

For SPRC~069, assuming the atomic gas is aligned with the optical disk, it yields an unphysically high rotation velocity of 1127$\pm$913~km~s$^{-1}$. If instead the gas resides predominantly in a perpendicular ring, the velocity is estimated as 214$\pm$13~km~s$^{-1}$, providing a plausible lower limit. The true rotation velocity likely lies between these extremes; both values are listed in columns (10)$-$(11) of Table~\ref{tab:co-add}.

Projection effects, as discussed in \citet{Deg2023MNRAS.525.4663D}, indicate that edge-on rings are readily identifiable in both images and velocity fields. Visual inspection of the PRG optical images confirms that the ring structures are typically edge-on. Comparing rotation velocities derived from the optical inclination with those assuming a 90$^\circ$ inclination allows us to infer whether the \HI\ gas is primarily associated with the central stellar body or the polar ring.

\subsection{Non-detection and Absorption}

The \HI\ mass upper limit is calculated following the methodology of \HI-MaNGA \citep{Masters2019MNRAS.488.3396M, Stark2021AAS...23752707S}, which assumes a line width of 200 km s$^{-1}$ and that the true flux is below the level corresponding to a 3$\sigma$ detection. The upper limit is given by:
\begin{multline}
M_{\rm H\ I, lim}/M_{\odot} = 2.356\times10^5 \times
\left(\frac{D_L}{\rm Mpc}\right)^2 \times \\
\left(\sqrt{200 \times (\Delta V/\rm km\ s^{-1})} \times \frac{3\sigma}{1000\ \rm mJy}\right),  
\label{equ:upper}
\end{multline}
\noindent where $D_L$ is the luminosity distance in Mpc, $\Delta V$ is the channel width of the rebinned \HI\ spectrum, and $\sigma$ is the corresponding noise level. For our sample, the spectra typically have a channel width of $\sim$6 km s$^{-1}$ and noise levels of 1.0$-$2.4 mJy. The resulting \HI\ mass upper limits for the non-detections range from $10^{8.4}$ M$_\odot$ to $10^{9.0}$ M$_\odot$.

Among the 20 new FAST \HI\ observations, one galaxy exhibits a potential absorption feature. SPRC~039 has a massive nearby companion, UGC~8225, within the FAST beam (Figure~\ref{fig:absorption}), and its \HI\ spectrum shows a 6.5$\sigma$ absorption signal. Given the spatial proximity and similar systemic velocities of the two galaxies, it is difficult to unambiguously identify which galaxy is the host of the \HI\ absorption.

\begin{figure*}
\epsscale{1.15}
\plotone{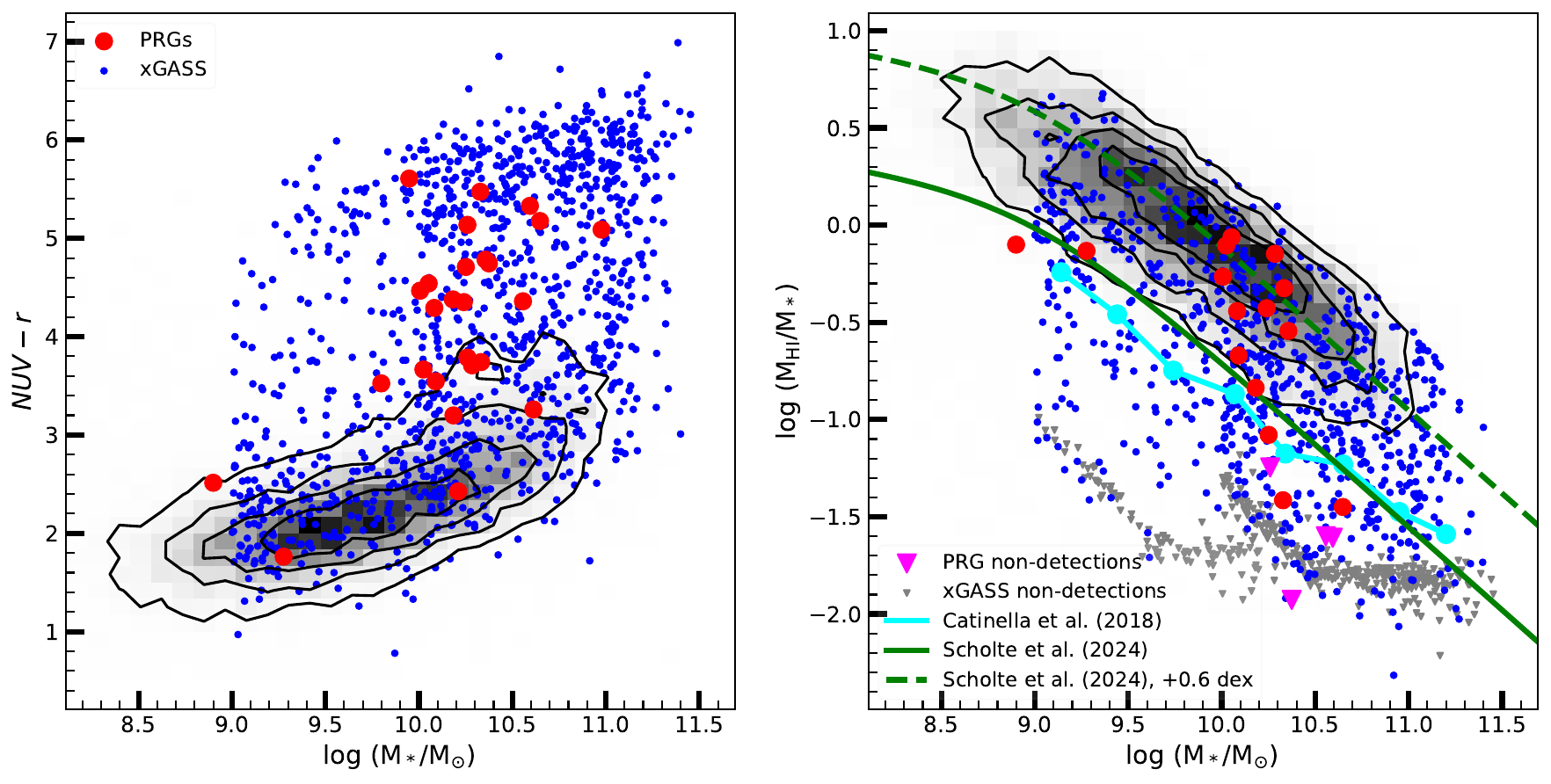}
\caption{
The NUV$-$r color (left) and atomic gas fraction (log~\MHI/\mstar) as a function of stellar mass. In each panel, the black contours show the distribution of ALFALFA-NSA galaxies, the blue symbols are xGASS galaxies, and the red circles are kinematically confirmed PRGs. In the right panel, down-triangles (navy: xGASS, magenta: PRGs) are \HI\ mass upper limits of non-detections, and the cyan curve shows the weighted average distribution of the xGASS galaxies from \citet{Catinella2018MNRAS.476..875C}. The green solid and dashed curves illustrate the relationship described by \citet{Scholte2024MNRAS.535.2341S}, as well as the 0.6~dex upper shift.
}
\label{fig:gf}
\end{figure*}

\subsection{\HI\ contents and distribution}
Among 40 kinematically confirmed PRGs, 27 have \HI\ information, including of 22 detections, 1 potential absorption, and 4 non-detections. 
We compare the NUV$-$r color and atomic gas fraction, $\log(M_{\rm HI}/M_\star)$, of PRGs as integrated systems with those of galaxies in the ALFALFA-NSA and xGASS surveys \citep{Catinella2018MNRAS.476..875C}. Relative to the \HI-detected ALFALFA-NSA galaxies, PRGs are redder and comparatively gas-poor (Figure~\ref{fig:gf}). 

However, compared to the unbiased, gas-fraction-limited xGASS sample, PRGs have higher atomic gas fractions, with only one-third of sources falling below the average relation.
xGASS is a gas fraction-limited census of the \HI\ content of approximately 1,  200 galaxies in the local Universe, selected solely on the basis of stellar mass (\mstar=10$^{9.0-11.5}$ \msun) and redshift (0.01$<$ z $<$0.05). The xGASS survey, a combination of GASS (\mstar =10$^{10.0-11.5}$) and GASS-low (\mstar =10$^{9.0-10.0}$), is designed to detect galaxies with a gas fraction that is limited to $\sim$2\%, otherwise classified as non-detections. The distribution of the upper limits in the right panel of Figure~\ref{fig:gf} is primarily attributable to the minor discrepancy in the gas fraction limit for GASS and GASS-low. It is evident that the PRG sample under consideration contains only six galaxies that fall below the weighted average curve delineated in \citet{Catinella2018MNRAS.476..875C}, given that the majority of them are massive galaxies.

But PRGs identified in the TNG50 Simulation are scattered in the Green Valley and star-forming main sequence \citep{Lopez-Castillo2025MNRAS.tmp.1846L}, even though most of them are classified as early-type galaxies. The inconsistency is mainly caused by the sample selection effects: nearly all PRGs in TNG50 have abundant gas and star formation in the ring structure. But our sample is a collection of all kinematically confirmed PRGs.

\begin{figure*}
\epsscale{1.1}
\plotone{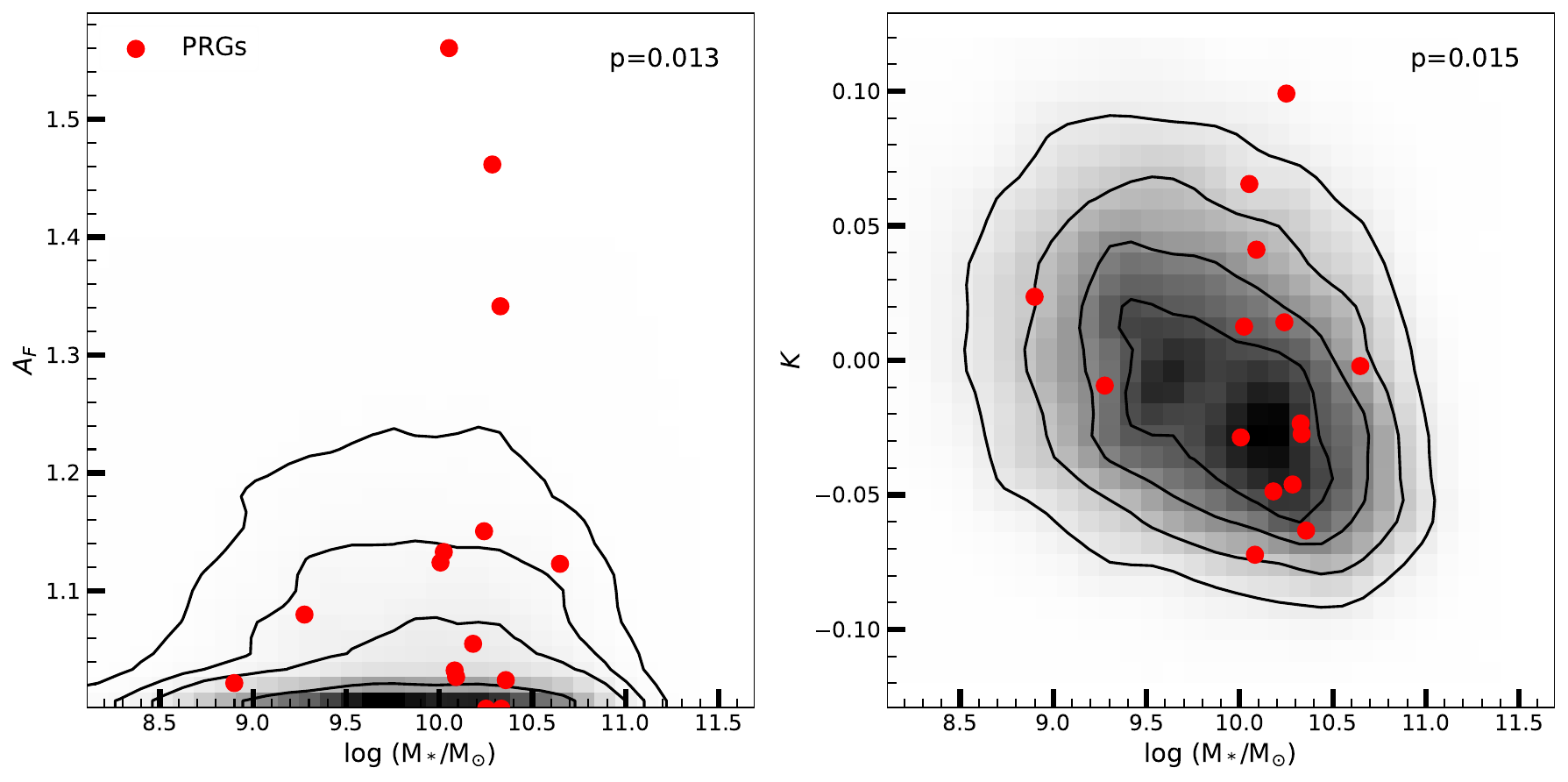}
\caption{
The \HI\ profile asymmetry ($A_F$, left) and profile shape ($K$, right) as a function of stellar mass for the ALFALFA-NSA sample and PRGs. In each panel, the black contours show the 20\%, 40\%, 60\%, and 80\% distribution of ALFALFA-NSA galaxies from inside out, and the red circles are kinematically confirmed PRGs. 
}
\label{fig:hi-distribution}
\end{figure*}

Besides the global NUV$-$r color and atomic gas fraction, we compare the \HI\ asymmetry and profile shape of PRGs with that of the ALFALFA-NSA sample (Figure~\ref{fig:hi-distribution}). The 2D Kolmogorov--Smirnov test is performed between the PRGs and the ALFALFA-NSA sample: $p<$0.05 indicates that the PRGs show significantly different distribution in the plane of log~\mstar\ versus asymmetry or shape. There are three galaxies (PRC~A-06, SPRC~033, and SPRC~260) showing significant \HI\ asymmetry ($A_F\geq$1.24, 3$\sigma$ outlier) among 15 PRGs. A \HI\ asymmetry fraction of 20\% is consistent with the ALFALFA sample \citep{Yu2022ApJS..261...21Y}. We caution that these three outlier have very high S/N \HI\ profiles, even though the asymmetry parameter itself depends on the observational effects and the definition \citep{Deg2020MNRAS.495.1984D}. Around 80\% galaxies show symmetric \HI\ profiles and simulations also show that the ring structure can be stable on a long timescale, i.e. $\geq$1~Gyr \citep{Smirnov2024MNRAS.527.4112S}. 

PRGs have slightly more single-peaked profiles (high $K$ values) than that of the ALFALFA-NSA sample. \citet{Zhang2025ApJ...993..149Z} find that the \HI\ profiles of H$\alpha$-star misaligned galaxies tend to be more single-peaked. Most of our galaxies show misaligned rotation for H$\alpha$ and star (see \citealt{Moiseev2011MNRAS.418..244M}), but the enhancement in the fraction of single-peaked profile is subtle. The consistency suggests external gas accretion may be the major formation mechanism for PRGs and misaligned galaxies.

Thus, PRGs predominantly occupy the green valley or are quenched, yet retain slightly elevated atomic gas fractions relative to the general galaxy population. There are slight differences on the \HI\ distribution based on the comparison of \HI\ profile asymmetry and profile shape for PRGs and the ALFALFA-NSA sample.

\begin{deluxetable}{crrrr}
\tablenum{3}
\tablecaption{The Dependence of $q_0$ on Stellar Mass and Redshift\label{tab:ba}}
\tablewidth{0pt}
\tabletypesize{\footnotesize}
\tablehead{
\colhead{$z$} &
\colhead{log $(M_{*}/M_{\odot})$} &
\colhead{$a$} &
\colhead{$b$} &
\colhead{$c$}
}
\startdata
$z<0.02$              & 7.6--10.8 & 0.03  & -0.55 &  2.75 \\
$0.02\leq z<0.04$     & 8.2--11.0 & 0.04  & -0.78 &  3.95 \\
$0.04\leq z<0.06$     & 8.5--11.2 & 0.04  & -0.77 &  3.97 \\
$0.06\leq z<0.08$     & 8.7--11.3 & 0.02  & -0.36 &  1.91 \\
$0.08\leq z<0.10$     & 8.9--11.3 & 0.01  & -0.24 &  1.24 \\
$0.10\leq z<0.12$     & 9.0--11.4 & -0.01 &  0.22 & -1.05 \\
$0.12\leq z<0.15$     & 9.1--11.5 & -0.02 &  0.54 & -2.71 \\
\enddata
\tablenotetext{a}{
Column (1): the range of redshift. Column (2): the range of stellar mass. Columns (3)--(5): parameters quantifying relations between $q_0$ and log~$(M_{*}$: $q_0$ = a$\times$ (log $(M_{*}/M_{\odot})$)$^2$+b$\times$ log $(M_{*}/M_{\odot})$+c.
}
\end{deluxetable}

\section{Discussions: The Tully--Fisher relation}
\label{sec:TFR}

\begin{figure*}
\epsscale{1.1}
\plotone{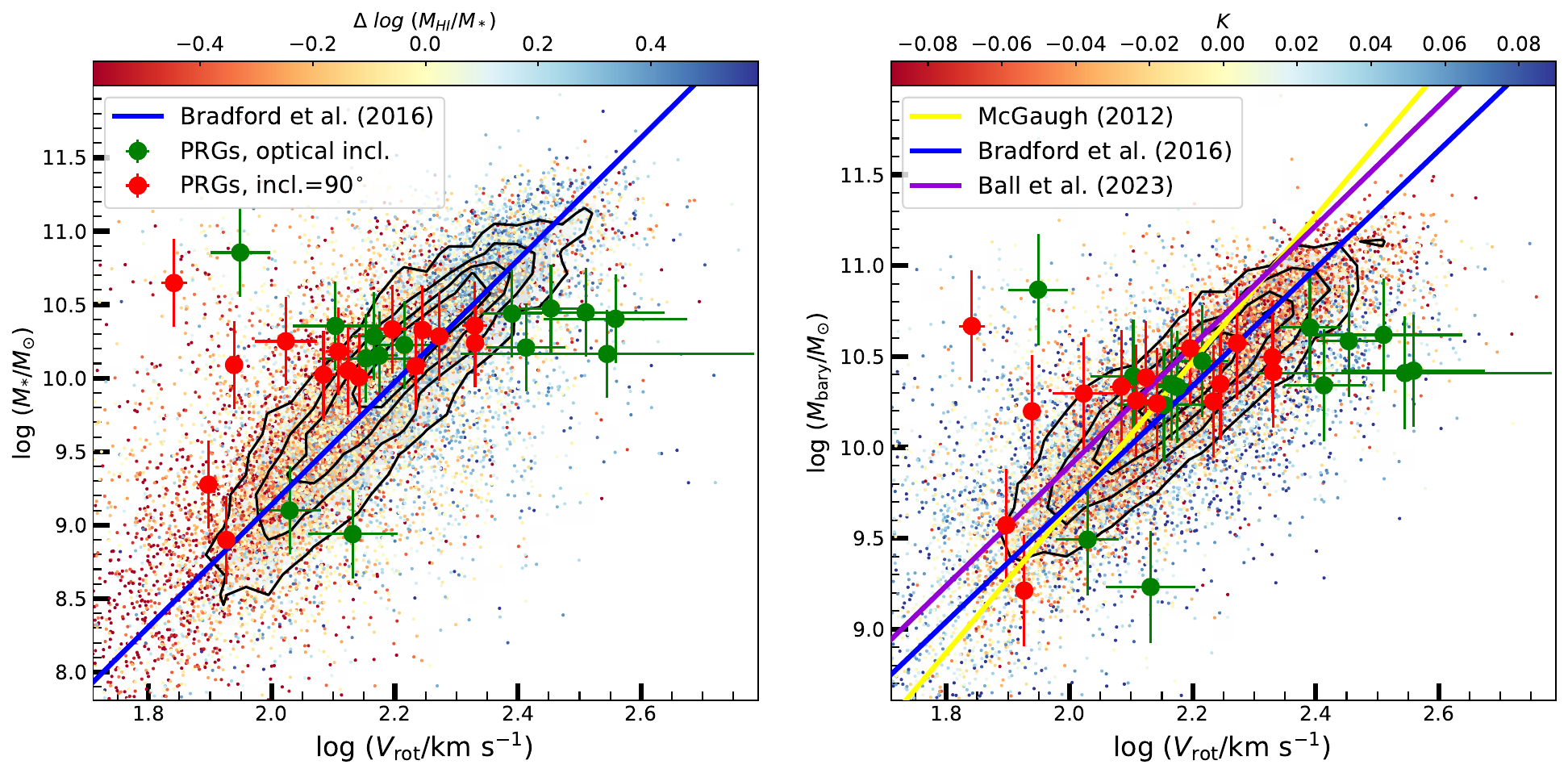}
\caption{
The stellar TFR (left) and baryoinc TFR (right) for the control sample and PRGs. The green (assuming an optical inclination angle) and red (assuming a 90$^{\circ}$ inclination angle) circles in each panels show the distribution of PRGs. The color-coded dots are the control sample before cutting the \HI\ profile shape and gas fraction deviation. The left panel is color-coded by the deviation of atomic gas fraction as a function of stellar mass, and the right panel is color-coded by the \HI\ profile shape $K$. The black contours in each panel show the levels of sample inclusion for the control sample, which are 20\%, 40\%, 60\%, and 80\% from inside out. The magenta line is the linear fitting result for the control sample after excluding 20\% extreme outliers. The blue, yellow, and darkviolet lines are TFRs from \citet{Bradford2016ApJ...832...11B}, \citet{McGaugh2012AJ....143...40M}, and \citet{Ball2023ApJ...950...87B}, respectively. 
}
\label{fig:tfr-prg}
\end{figure*}

The distributions of PRGs in the TFR and bTFR is investigated and compared with that of the control sample (Figure~\ref{fig:tfr-prg}). The TFRs of PRGs demonstrate a considerable degree of variability, irrespective of the assumption of a $90^\circ$ inclination angle or the utilisation of the optical inclination angle. A 2D Kolmogorov--Smirnov test is employed to quantitatively assess the differences in TFRs for PRGs and controls. The results obtained, $p\lesssim$0.02, indicate that both the TFR and bTFR for PRGs--whether corrected using the optical inclination or assuming $90^\circ$--differ significantly from those of control galaxies. For those well-defined samples, a smaller scatter in TFRs is consistently observed \citep{ McGaugh2012AJ....143...40M, Bradford2016ApJ...832...11B, Ball2023ApJ...950...87B}. Furthermore, the ring structure is faint relative to the central component, but a correction of $\sim10-$30\% \citep{Lackey2024Galax..12...42L} is insufficient to fully account for the observed TFR offsets. 

Therefore, PRGs do not follow the TFR like most rotation-dominated galaxies. At a given stellar mass, PRGs exhibit higher rotation velocities than the controls if their \HI\ gas is assumed to be aligned with the optical disk, same hypothesis and results with literature studies \citep{vanDriel2002A&A...386..140V, Iodice2003ApJ...585..730I}. While assuming a $90^\circ$ inclination angle, the distribution of PRGs in TFRs is found to be more consistent with that of the control sample, even though a small number of outliers show decreased rotation velocity at a given stellar or baryonic mass. In fact, \citet{Deg2023MNRAS.525.4663D} find that approximately half of the H I gas resides in the ring in two PRGs. This indicates \HI\ gas accretion in these systems was quite significant, possibly causing the dispersion in TFR for PRGs. Comparing the distribution of PRGs under two inclination angle assumption, we expect that most \HI\ gas is distributed in the perpendicular ring in our PRGs sample. But more spatially resolved observations are required to determine the gas distribution and kinematics in PRGs robustly.

However, the majority of PRGs are characterised by a relatively narrow stellar mass range, typically spanning from $10^{10.0}$ to $10^{10.5}$~M$_\odot$. It is notable that their rotation velocities exhibit a substantial range. If we ignore the two galaxies with a stellar mass smaller than 10$^{9.5}$~\msun, all trends for PRGs are quite flat in TFRs. The significant variation suggests that the gas in PRGs is not rotating coherently: the gas in the ring does not necessarily form the same rotation pattern as that in the main body, if indeed there is any such pattern.

To further understand the distribution and scatter of PRGs in TFRs, we color-coded the TFR and bTFR for the ALFALFA-NSA sample as the deviation of the atomic gas fraction and \HI\ profile shape. Figure~\ref{fig:tfr-prg} shows that gas-deficient galaxies or those galaxies exhibiting single-peaked \HI\ profiles (positive $K$ values) contribute significantly to the observed large scatter in TFRs for controls. Outliers located in the upper-left region of the TFR diagram are likely to be affected by a depletion of atomic gas in the optical disk. In these systems, the majority of the cold gas is concentrated in the central region, resulting in a significant underestimation of the rotation velocity inferred from the \HI\ line width. Conversely, galaxies located in the lower-right region have accreted substantial gas, frequently exhibiting disturbed kinematics, leading to an overestimation of rotation velocities based on the \HI\ line width.

We select galaxies with double-horned or flat-topped \HI\ profiles and a high gas fraction deviation (see Section~\ref{subsec:control}), and demonstrate the 20\%, 40\%, 60\%, and 80\% percent distribution as black contours in Figure~\ref{fig:tfr-prg}. It is acknowledged that the exclusion of galaxies exhibiting single-peaked H I profiles inherently precludes the consideration of low-mass galaxies. Nevertheless, these galaxies are found to be significantly influenced by mergers and radial motions, thereby rendering them inadequate for the reliable estimation of rotation measures \citep{Downing2023MNRAS.522.3318D}. At a fixed optical inclination, stellar mass, optical concentration, and \HI\ mass, galaxies with higher global or inner star formation rates tend to have more single-peaked profiles \citep{Yu2022ApJ...930...85Y}, consistent with a gas-fueling scenario. In addition, massive galaxy mergers exhibit a higher frequency of single-peaked profiles in comparison with control galaxies \citep{Zuo2022ApJ...929...15Z}.

The TFR for controls, the ALFALFA-NSA sample, shows a slight turn-over trend at the high-mass end, i.e., log~(\mstar/\msun)$>$10.5. As \citet{Peletier1993ApJ...418..626P} and \citet{Noordermeer2007MNRAS.381.1463N} have demonstrated, analogous trends have been observed. Galaxies with $10.8 \lesssim \log M_{\rm bary} \lesssim 11.8$ and $0.2 < z < 0.4$ cover a wide range of rotation velocities \citep{Catinella2015MNRAS.446.3526C, Jarvis2025arXiv250611935J}, indicating that gas kinematics can vary significantly at the high-mass end. It has been demonstrated that massive early-type galaxies frequently manifest unsettled gas distributions or surrounding gas clouds \citep{Serra2012MNRAS.422.1835S}, which can result in an overestimation of rotation velocity from integrated \HI\ spectra. The narrow stellar mass range and the wide range of rotation velocity indicate that these kinematically confirmed PRGs may have more disturbed gas distribution or kinematics than rotation-dominated galaxies.

It can thus be concluded that the primary cause of the significant dispersion of TFRs is the imperfect probing of the galaxy's rotation by the H I distribution or velocity field. The disparities in the TFR for PRGs may be attributed to a number of factors: (1) the atomic gas has not yet been virialized, leading to the rotation velocity overestimation; (2) misalignment between the optical disk and \HI\ disk due to substantial gas accretion; and (3) significant uncertainties in stellar mass measurements caused by irregular optical morphology or complex star formation history, particularly the presence of extended optical rings. 

However, it must be noted that the current sample size of the PRG is insufficient to draw definitive conclusions. The discrepancy between TFRs derived under different inclination assumptions underscores the significance of the \HI\ spatial distribution for comprehending the intrinsic properties of PRGs.

\section{Summary}
\label{sec:sum}
We compile all kinematically confirmed PRGs from the literature and compare their gas content, NUV$-$r colors, and rotation velocities with that of controls. The new FAST tracking and MultibeamOTF observations supplement \HI\ data for collected PRGs. The \HI\ line widths and fluxes are measured using the curve-of-growth method, thereby providing constraints on atomic gas content and rotation. A control sample is constructed by cross-matching the ALFALFA survey \citep{Haynes2011AJ....142..170H, Haynes2018ApJ...861...49H} with the NSA catalogue \citep{Blanton2011AJ....142...31B}. This allows for a comparison of PRG properties with those of typical galaxies.

\begin{description}
\item[$\bullet$ A catalog of kinematically confirmed PRGs.] 
This sample represents the most comprehensive and complete collection of kinematically confirmed polar ring galaxies available so far.
Pronounced kinematic misalignment between the major and minor axes has been observed in all 40 PRGs. Combining new FAST observations with literature data, we obtain 22 \HI\ detections, 1 potential \HI\ absorption, and 4 non-detections. For detections, we measure their \HI\ profile asymmetry, shape, \HI\ mass, and rotation velocity.

\item[$\bullet$ Atomic gas and star formation.] PRGs primarily inhabit the green valley or quenched population. But they are slightly more gas-rich than galaxies in the xGASS survey, which is a gas fraction-limited census of the \HI\ content in local Universe. It is hypothesised that the elevated atomic gas fraction is the result of gas accretion forming the polar ring. The red NUV$-$r thus low star formation activity suggests that \HI\ does not fuel star formation efficiently. PRGs exhibit slightly more asymmetric and single-peaked \HI\ profiles than that of the ALFALFA-NSA sample. 

\item[$\bullet$ PRGs demonstrate a large scatter in the TFRs.] Assuming either a $90^\circ$ inclination or an optical inclination angle, PRGs exhibit a significant scatter in the TFR and bTFR. If the gas is assumed aligned with the optical disk, PRGs exhibit higher rotation velocities at a given stellar or baryonic mass. It is consistent with literature studies. When we assume a $90^\circ$ inclination, the scatter decrease substantially, but still noticeable. If we assume PRGs follow the TFR distribution, the majority of atomic gas in PRGs should be located in the polar ring.

\end{description}

In conclusion, PRGs (1) are mainly green-valley or quenched galaxies; (2) exhibit a higher gas richness compared to their massive early-type counterparts, (3) have similar \HI\ profile asymmetry and shape than the \HI-detected galaxies, and (4) have their \HI\ gas mainly distributed in the polar ring, instead of the main stellar body. The results of this study indicate that accreted gas requires a considerable timescale to settle, virialize, and fuel star formation.

\begin{acknowledgments}
We thank the anonymous referee for helpful comments and suggestions. 
This work was supported by the National Science Foundation of China No. 12588202, 11973051, 12041302, 12373012, 12473023, and U1931110, the China Postdoctoral Science Foundation (2022M723175, GZB20230766), the National SKA Program of China (No. 2025SKA0130100), the International Partnership Program of Chinese Academy of Sciences, Program No. 114A11KYSB20210010, National Key\&D Program of China No. 2023YFE0110500, 2023YFA1608004, and 2023YFC2206403, the National Natural Science Foundation of China No. 11903003, and the Ministry of Science and Technology of China No. 2022YFA1605300. T.C.W. is supported by CAS project No. JZHKYPT-2021-06. LCH was supported by the National Science Foundation of China (12233001) and the China Manned Space Program (CMS-CSST-2025-A09). N.T. acknowledges the support by the University Annual Scientific Research Plan of Anhui Province (Nos. 2023AH030052). This work is also supported by the Young Researcher Grant of the Institutional Center for Shared Technologies and Facilities of the National Astronomical Observatories, Chinese Academy of Sciences. 
This work made use of the data from FAST (Five-hundred-meter Aperture Spherical radio Telescope). FAST is a Chinese national mega-science facility, operated by the Nathaiional Astronomical Observatories, Chinese Academy of Sciences.
This research made use of the NASA/IPAC Extragalactic Database ({\url http://ned.ipac.caltech.edu}), which is funded by the National Aeronautics and Space Administration and operated by the California Institute of Technology. We used Astropy, a community-developed core Python package for astronomy \citep{Astropy2013A&A...558A..33A, Astropy2018AJ....156..123A}. We perform the 2D Kolmogorov--Smirnov test using the public code \textsc{ndtest}\footnote{Written by Zhaozhou Li, \url{https://github.com/syrte/ndtest}}.
\end{acknowledgments}

\vspace{5mm}
\facilities{FAST}

\software{astropy \citep{Astropy2013A&A...558A..33A, Astropy2018AJ....156..123A}
          }

\vspace{5mm}
\bibliography{}{}
\bibliographystyle{aasjournal}

\end{document}